\documentclass[conference]{IEEEtran}

\IEEEoverridecommandlockouts
\usepackage{cite}
\usepackage{amsmath,amssymb,amsfonts}
\usepackage{algorithm}
\usepackage{algorithmic}
\usepackage{graphicx}
\usepackage{textcomp}
\usepackage{xcolor}
\usepackage{comment}
\usepackage{float}
\usepackage{booktabs}
\usepackage{multirow}
\usepackage[hyphens]{url}
\usepackage[hidelinks]{hyperref}
\begin{document}

\title{A Persistent-State Dataflow Accelerator for Memory-Bound Linear Attention Decode on FPGA}

\author{\IEEEauthorblockN{Neelesh Gupta\textsuperscript{*}, Peter Wang\textsuperscript{*}, Rajgopal Kannan\textsuperscript{\dag} and Viktor K. Prasanna\textsuperscript{*}}

\IEEEauthorblockA{\textsuperscript{*}University of Southern California, USA\\
\textsuperscript{\dag}DEVCOM Army Research Office, USA\\
Email: \{neeleshg, plwang, prasanna\}@usc.edu, rajgopal.kannan.civ@army.mil}
}
\maketitle

% ================================================================
% \begin{abstract}
% Gated DeltaNet (GDN) is a linear attention mechanism that replaces
% the growing KV cache with a fixed-size recurrent state.
% Hybrid LLMs like Qwen3-Next use 75\% GDN layers and achieve
% competitive accuracy to attention-only models. However, at batch-1, GDN decode is severely memory-bound on GPUs since the
% full recurrent state must be round-tripped through HBM every token. We show that this bottleneck is
% architectural, not algorithmic: all subquadratic sequence models exhibit
% arithmetic intensities below 1\,FLOP/B at decode time, making them
% \emph{more} memory-bound than standard Transformers.
% We present an FPGA accelerator that eliminates this bottleneck by holding
% the full 2\,MB recurrent state persistently in on-chip BRAM, converting
% the workload from memory-bound to compute-bound. Our design fuses the
% GDN recurrence into a five-phase pipelined datapath that performs only
% one read and one write pass over each state matrix per token, exploits
% Grouped Value Attention for paired-head parallelism, and overlaps
% preparation, computation, and output storage via dataflow pipelining.
% We explore four design points on an AMD Alveo U55C using Vitis HLS,
% varying head-level parallelism from 2 to 16 value-heads per iteration.
% Our fastest configuration achieves 63\,$\mu$s per token, 4.5$\times$
% faster than the GPU reference implementation on an NVIDIA H100 PCIe.
% Post-implementation power analysis measures 9.96\,W on-chip,
% yielding up to 60$\times$ energy efficiency.
% \end{abstract}
\begin{abstract}
Gated DeltaNet (GDN) is a linear attention mechanism that replaces
the growing KV cache with a fixed-size recurrent state.
Hybrid LLMs like Qwen3-Next use 75\% GDN layers and achieve
competitive accuracy to attention-only models. However, at batch-1, GDN decode is memory-bound on GPUs since the
full recurrent state must be round-tripped through HBM every token. We show that this bottleneck is
architectural, not algorithmic, as all subquadratic sequence models exhibit
arithmetic intensities below 1\,FLOP/B at decode time, making them
\emph{more} memory-bound than standard Transformers.
We present an FPGA accelerator that eliminates this bottleneck by holding
the full 2\,MB recurrent state persistently in on-chip BRAM, converting
the workload from memory-bound to compute-bound. Our design fuses the
GDN recurrence into a five-phase pipelined datapath that performs only
one read and one write pass over each state matrix per token, exploits
Grouped Value Attention for paired-head parallelism, and overlaps
preparation, computation, and output storage via dataflow pipelining.
We explore four design points on an AMD Alveo U55C using Vitis HLS,
varying head-level parallelism from 2 to 16 value-heads per iteration.
Our fastest configuration achieves 63\,$\mu$s per token, 4.5$\times$
faster than the GPU reference on NVIDIA H100 PCIe.
Post-implementation power analysis reports 9.96\,W on-chip, yielding up to
60$\times$ greater energy efficiency per token decoded.
% Post-implementation power reports of the placed design measure 9.96\,W on-chip with up to 60$\times$ energy efficiency per token decoded.
\end{abstract}


% ================================================================
\section{Introduction}

Large language models (LLMs) have become central to a wide range of
applications, from conversational assistants to code generation and scientific
reasoning. However, their deployment is constrained by the cost of inference,
which is dominated by two bottlenecks: the arithmetic cost of matrix
multiplications and the memory cost of maintaining per-layer context. In
standard Transformer architectures~\cite{vaswani2017attention}, the key-value (KV) cache grows linearly
with sequence length, consuming increasing amounts of accelerator memory as
conversations or documents grow longer.

\begin{figure}[t]
  \centering
  \includegraphics[width=\columnwidth]{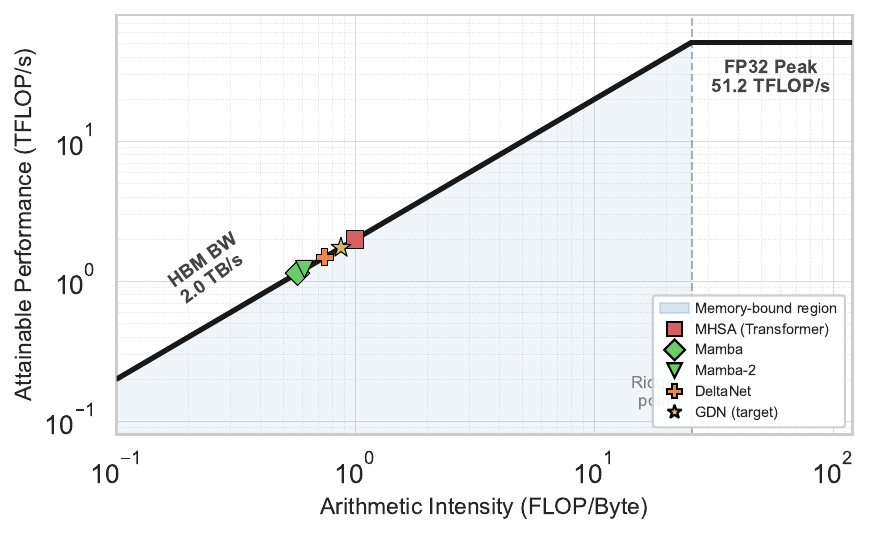}
  % \caption{Batch-1 decode arithmetic intensity on the H100 FP32 roofline for LLM decode is memory-bound where
  % subquadratic models
  % (GDN~\cite{yang2025gated},
  %  DeltaNet~\cite{yang2024deltanet},
  %  Mamba~\cite{gu2024mamba},
  %  Mamba-2~\cite{dao2024mamba2})
  % are \emph{more}
  % memory-bound than multi-head softmax attention (MHSA) with arithmetic intensities well below 1\,FLOP/B.}

  \caption{Batch-1 decode arithmetic intensity on the H100 FP32 roofline.                                                                                
    All architectures fall deep in the memory-bound regime.                                                                                              
    MHSA Transformers~\cite{vaswani2017attention} achieve an arithmetic                                                                                  
    intensity near 1\,FLOP/B whereas subquadratic models                                                                                               
    (GDN~\cite{yang2025gated},                                                                                                                           
     DeltaNet~\cite{yang2024deltanet},                                                                                                                   
     Mamba~\cite{gu2024mamba},
     Mamba-2~\cite{dao2024mamba2})
    fall well below 1\,FLOP/B.}
  \label{fig:roofline}
\end{figure}

To address these scaling limitations, subquadratic sequence
models~\cite{gu2024mamba,dao2024mamba2,yang2024deltanet} replace softmax
attention with fixed-size recurrent state, reducing per-layer memory from
$O(n)$ to $O(1)$ in sequence length. Gated DeltaNet
(GDN)~\cite{yang2025gated} combines selective gating with an
error-correcting delta rule and has been adopted by production-scale
architectures: Qwen3-Next~\cite{qwen2025qwen3next} stacks three GDN layers for every one
softmax attention layer, making GDN decode the dominant per-token
primitive.

Critically, while LLM decode is broadly
memory-bound~\cite{shazeer2019fast,pope2023efficiently,leviathan2023fast,yuan2024llm,kim2024squeezellm,jayanth2024benchmarking,he2025lutllm,he2025waferllm},
subquadratic architectures are \emph{more} so than standard attention
(Figure~\ref{fig:roofline}). MHSA decode with grouped-query attention (GQA) achieves an
arithmetic intensity near 1\,FLOP/B, whereas recurrent models like GDN
fall below: each layer must read and write its \emph{entire} fixed-size
state matrix every token, yielding only $\sim$0.87\,FLOP/B---well below
the H100's ridge point of 25.6\,FLOP/B. Optimized
kernels~\cite{flashinfer2025} reduce software overhead but cannot
eliminate the fundamental HBM round-trip.

We observe that the full 2\,MB GDN state fits in modern FPGA on-chip
BRAM. By holding it persistently across kernel invocations, we eliminate
the off-chip bottleneck entirely, converting the workload from
memory-bound to compute-bound. We present a dataflow FPGA accelerator
that exploits this insight.
Our contributions are as follows:
\begin{itemize}
\item We propose the first FPGA accelerator for autoregressive Gated
  DeltaNet decode, holding the full 2\,MB recurrent state persistently
  in on-chip BRAM to eliminate off-chip memory round-trips that
  limit batch-1 GPU performance.
\item We fuse the GDN recurrence into a five-phase tiled datapath that
  performs only one read pass and one write pass over each state matrix
  per token, halving the naive state-access cost through an algebraic
  restructuring of the output computation.
\item We exploit Grouped Value Attention structure to share query/key
  datapaths across pairs of values, scaling head-level parallelism
  without increasing pipeline interval.
\item We evaluate four configurations on an Alveo U55C, achieving up to
  4.5$\times$ lower latency than an H100 GPU. Post-implementation
  power analysis measures 9.96\,W on-chip, yielding up to
  60$\times$ greater energy efficiency.
\end{itemize}

% ================================================================
\section{Background}

\subsection{Qwen3-Next and Hybrid LLM Architectures}

\begin{figure}[t]
  \centering
  \includegraphics[width=\columnwidth]{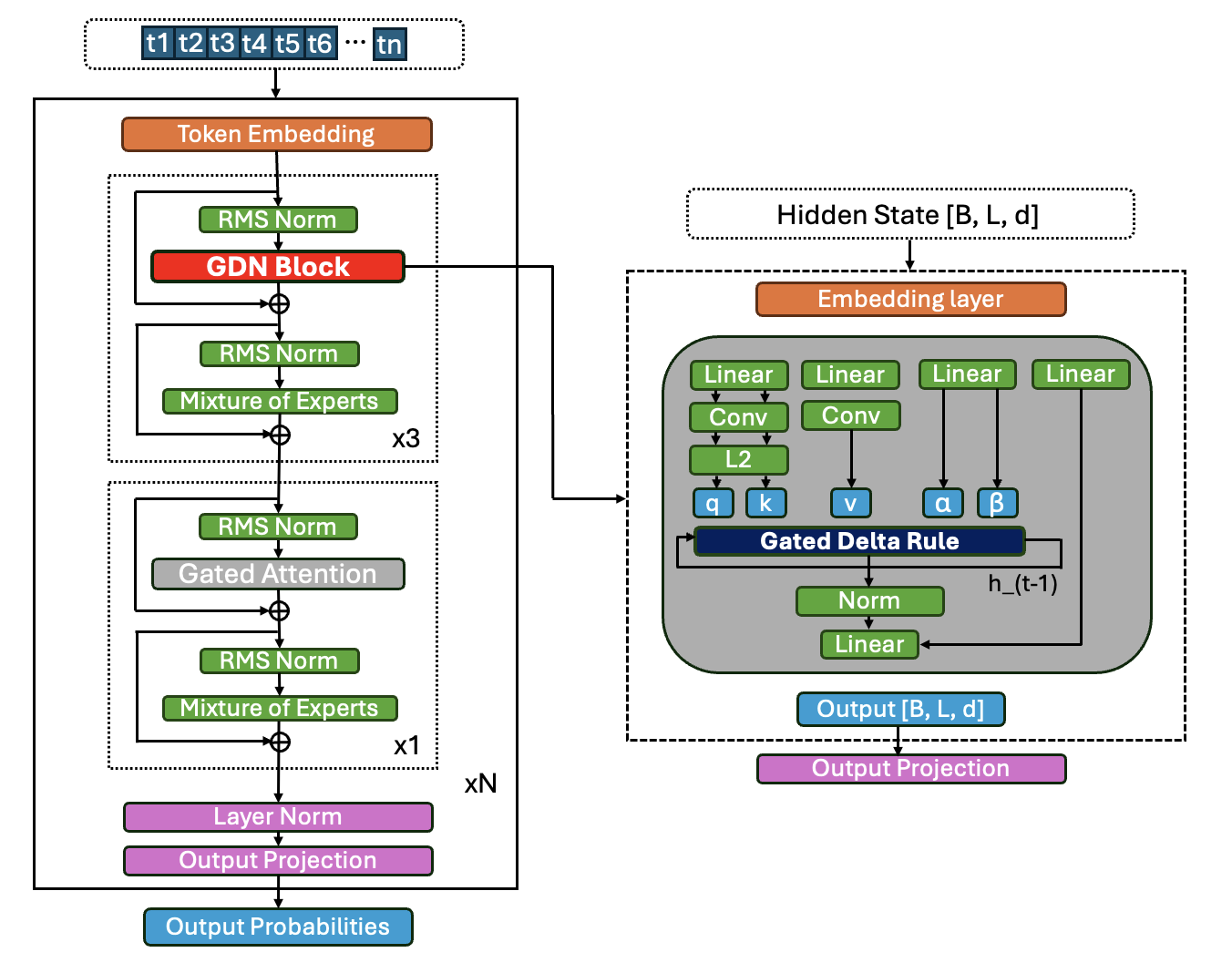}
  \caption{Qwen3-Next~\cite{qwen2025qwen3next} hybrid architecture. GDN layers (3:1 ratio)
  replace softmax attention with fixed-size recurrent state, making GDN
  decode the dominant per-token primitive.}
  \label{fig:qwen3}
\end{figure}

Next-generation LLMs increasingly adopt hybrid architectures that interleave
full softmax attention layers with sub-quadratic alternatives.
Qwen3-Next~\cite{qwen2025qwen3next} uses a 3:1 ratio of Gated
DeltaNet~\cite{yang2025gated} layers to full-attention layers
(Figure~\ref{fig:qwen3}). Each GDN layer replaces the growing KV-cache with
a fixed-size recurrent state matrix $S \in \mathbb{R}^{d \times d}$
($d{=}128$), reducing per-layer memory from $O(n)$ to $O(1)$ in sequence
length. In the Qwen3-Next configuration, each layer has $h_q = h_k = 16$
query/key heads and $h_v = 32$ value heads arranged in a 2:1 Grouped Value
Attention (GVA) ratio: every query/key pair serves two value-heads that
maintain independent state matrices but share the query and key vectors.

\subsection{Prefill vs.\ Decode}

LLM inference proceeds in two phases. During \emph{prefill}, the model
processes the full input prompt in parallel, and the GDN state matrices can
be computed via efficient chunkwise parallel
algorithms~\cite{yang2024deltanet}. During \emph{decode}, the model generates
one token at a time autoregressively. Each decode step must read the current
state, compute a single-token update, write the new state, and produce one
output vector per head. At batch~1 this workload is extremely small
($\sim$4.2\,M FLOPs) relative to the state that must be accessed
(32 matrices of $128 \times 128$ floats = 2\,MB). On GPU, this means decode
is dominated by the cost of moving state through the memory hierarchy rather
than by arithmetic---a classic memory-bandwidth bottleneck.

\subsection{Gated DeltaNet Recurrence}

Gated DeltaNet~\cite{yang2025gated} combines the selective gating of
Mamba2~\cite{dao2024mamba2} with the \emph{delta rule} from
DeltaNet~\cite{yang2024deltanet}. The delta rule originates from
associative memory: the state matrix $S$ acts as a key--value store where
$S^{\!\top} k$ retrieves the value currently associated with key $k$.
Rather than blindly overwriting this association, the delta rule computes
the prediction error $v - S^{\!\top}\!k$ and updates $S$ proportionally,
analogous to the Widrow--Hoff learning rule. This error-correcting
mechanism allows the model to update stored associations incrementally
without catastrophic overwriting.

For a single value-head $h$ with state
$S_{t-1} \in \mathbb{R}^{d \times d}$, the decode-step recurrence is:
\begin{align}
  r_t &= S_{t-1}^{\!\top} k_t
  & &\text{(retrieval)} \label{eq:retrieval} \\
  \Delta v_t &= \beta_t\,(v_t - r_t)
  & &\text{(delta correction)} \label{eq:delta} \\
  S_t &= g_t \cdot S_{t-1} + k_t \, \Delta v_t^{\!\top}
  & &\text{(state update)} \label{eq:update} \\
  o_t &= \tfrac{1}{\sqrt{d}}\; S_t^{\!\top} q_t
  & &\text{(output)} \label{eq:output}
\end{align}
where $q_t, k_t \in \mathbb{R}^d$ are query and key vectors,
$v_t \in \mathbb{R}^d$ is the value vector, and the scalar gates are:
\begin{align}
  g_t &= \exp\!\bigl(-\sigma(\alpha_t) \cdot e^{A_{\log}} \cdot
         \mathrm{softplus}(dt_{\mathrm{bias}})\bigr) \label{eq:gate_g} \\
  \beta_t &= \sigma(b_t) \label{eq:gate_beta}
\end{align}
Here $A_{\log}$ and $dt_{\mathrm{bias}}$ are learned per-head parameters,
while $\alpha_t$ and $b_t$ are token-dependent inputs.

% ================================================================
\section{Performance Model}

\subsection{Problem Formulation}

Table~\ref{tab:platforms} summarizes the two platforms we compare.
The key asymmetry is on-chip memory \emph{type}: the H100's 50\,MB
L2 is a hardware-managed cache with no guarantee of persistence
across kernel invocations, so the 2\,MB GDN state must be
round-tripped through HBM every token. The U55C's 17.6\,MB of BRAM,
by contrast, is software-managed scratchpad that retains data
indefinitely across invocations.

\begin{table}[t]
\centering
\caption{Platform Specifications}
\label{tab:platforms}
\begin{tabular}{lcc}
\toprule
& \textbf{NVIDIA H100 PCIe} & \textbf{AMD Alveo U55C} \\
\midrule
Technology & TSMC 4N & TSMC 16 nm \\
FP32 peak & 51\,TFLOP/s & $\sim$2.7\,TFLOP/s\textsuperscript{a} \\
Memory BW & 2.0\,TB/s (HBM3) & 460\,GB/s (HBM2e) \\
On-chip memory & 50\,MB (L2 cache) & 17.6\,MB (BRAM) \\
BRAM blocks & --- & 4{,}032 $\times$ BRAM\_18K \\
DSP slices & --- & 9{,}024 $\times$ DSP48E2 \\
Board TDP & 350\,W & 150\,W \\
\bottomrule
\multicolumn{3}{l}{\footnotesize \textsuperscript{a}$9{,}024 \text{ DSPs} \times 300\,\text{MHz}$; effective throughput is data\-path-dependent.}
\end{tabular}
\end{table}

A single GDN decode step must update $h_v{=}32$ persistent state
matrices $S^{(h)} \in \mathbb{R}^{d \times d}$ ($d{=}128$, FP32) and
produce an output vector from per-token inputs. The aggregate state
totals $h_v \cdot d^2 \cdot w = 2$\,MB ($w{=}4$ bytes for FP32),
while the per-token compute is only $\sim$4.2\,M FLOPs.
Our design objective is to minimize the per-token decode latency:
\begin{equation}
  \min_{H_{\mathrm{iter}}} \quad L \;=\;
    \frac{h_v}{H_{\mathrm{iter}}} \cdot T_{\mathrm{iter}}
    + T_{\mathrm{load}}
\label{eq:latency_obj}
\end{equation}
subject to the on-chip memory constraint that the full state fits in
BRAM and the FPGA fabric constraints that scale with parallelism:
\begin{align}
  h_v \cdot d^2 \cdot w &\;\leq\; B_{\mathrm{BRAM}}
  &
    % & &\text{(persistent state)}
    \label{eq:c_bram} \\
  R_k(H_{\mathrm{iter}}) &\;\leq\; R_k^{\max},\;\;
    k \in \{\text{DSP, FF, LUT}\}
    % & &\text{(fabric resources)}
    \label{eq:c_fabric}
\end{align}
Here $H_{\mathrm{iter}}$ is the number of value-heads processed per
dataflow iteration---the primary design knob. $T_{\mathrm{iter}}$ is
the per-iteration pipeline interval (dominated by two passes over a
$d \times d$ state matrix), and $T_{\mathrm{load}}$ is the one-time
cost of loading per-token inputs from host memory via AXI.

Constraint~\eqref{eq:c_bram} is the enabling precondition: if the
state fits on-chip, the off-chip memory bottleneck is eliminated
entirely.  On the U55C, $2\,\text{MB} \ll 17.6\,\text{MB}$, so this
holds comfortably.  Constraint~\eqref{eq:c_fabric} bounds the
compute resources that grow with $H_{\mathrm{iter}}$; as shown in
Section~\ref{sec:results}, the optimal $H_{\mathrm{iter}}{=}8$ stays
under 25\% for all resource types.

\subsection{Baseline Cost Analysis}

Algorithm~\ref{alg:naive} shows the standard sequential GDN decode
step for a single value-head~\cite{yang2025gated}. This na\"ive
approach requires \emph{three} full passes over the $d \times d$ state
matrix: one for retrieval (line~\ref{line:retrieve}), one for state
update (line~\ref{line:update}), and one for output
(line~\ref{line:output}).

\begin{algorithm}[t]
\caption{Standard GDN Decode Step (Single Head)}
\label{alg:naive}
\begin{algorithmic}[1]
\REQUIRE $q, k, v \in \mathbb{R}^d$, state $S \in \mathbb{R}^{d \times d}$, gates $g, \beta \in \mathbb{R}$
\ENSURE output $o \in \mathbb{R}^d$, updated state $S$
\STATE $r \leftarrow S^{\!\top} k$ \hfill\COMMENT{Retrieval: read pass 1} \label{line:retrieve}
\STATE $\Delta v \leftarrow \beta \cdot (v - r)$ \hfill\COMMENT{Delta correction}
\STATE $S \leftarrow g \cdot S + k \, \Delta v^{\!\top}$ \hfill\COMMENT{State update: read+write pass 2} \label{line:update}
\STATE $o \leftarrow \tfrac{1}{\sqrt{d}} \; S^{\!\top} q$ \hfill\COMMENT{Output: read pass 3} \label{line:output}
\end{algorithmic}
\end{algorithm}

On FPGA with column parallelism $P_K$, each pass over a $d \times d$
matrix costs $d \times (d / P_K)$ cycles at initiation interval~1. With
$d{=}128$ and $P_K{=}16$, this gives 1{,}024 cycles per pass.
The na\"ive per-head cost is therefore:
\begin{equation}
  T_{\mathrm{head}}^{\mathrm{naive}} = 3 \times \frac{d^2}{P_K}
  = 3{,}072 \text{ cycles}
  \label{eq:naive_cost}
\end{equation}
Processing all $h_v$ heads sequentially in groups of
$H_{\mathrm{iter}}$ yields the na\"ive total:
$L^{\mathrm{naive}} = (h_v / H_{\mathrm{iter}}) \times 3{,}072
+ T_{\mathrm{load}}$.

\subsection{Latency Decomposition}

We decompose $T_{\mathrm{iter}}$ into the sum of per-phase costs within
the compute pipeline. The two state-access phases (read and write) each
cost $d^2 / P_K$ cycles; the remaining phases (dot product, delta
correction, output) each cost $d / P_K$ cycles and are negligible:
\begin{equation}
  T_{\mathrm{iter}} \approx 2 \times \frac{d^2}{P_K}
  + 3 \times \frac{d}{P_K}
  = 2{,}048 + 24 = 2{,}072 \text{ cycles}
  \label{eq:titer}
\end{equation}
In practice, pipeline startup and dataflow scheduling add a small
overhead, giving $T_{\mathrm{iter}} \approx 2{,}106$ cycles (verified by
HLS reports). The total latency model is then:
\begin{equation}
  L = \frac{h_v}{H_{\mathrm{iter}}} \times 2{,}106 + T_{\mathrm{load}}
  \label{eq:latency_model}
\end{equation}
This reveals three levers for reducing latency: (1)~eliminate one of the
state-access passes to reduce $T_{\mathrm{iter}}$; (2)~increase
$H_{\mathrm{iter}}$ to reduce the number of iterations; and
(3)~overlap stages within each iteration via pipelining.
Table~\ref{tab:profile} shows the precondition enabling all three: if
the state fits on-chip, state I/O vanishes, making the workload
compute-bound.

\begin{table}[t]
\centering
\caption{Per-Token Computational Profile ($h_v{=}32$, $d{=}128$, FP32)}
\label{tab:profile}
\begin{tabular}{lcc}
\toprule
\textbf{Quantity} & \textbf{GPU} & \textbf{FPGA (ours)} \\
\midrule
Compute (FLOPs) & \multicolumn{2}{c}{$\sim$4.2\,M} \\
State I/O (bytes) & 4,194,304 & 0 \\
Token I/O (bytes) & 49,664 & 49,664 \\
\midrule
Total off-chip I/O & $\sim$4.24\,MB & $\sim$48.5\,KB \\
Op.\ intensity (FLOP/B) & $\sim$1.0 & $\sim$88 \\
\bottomrule
\end{tabular}
\end{table}

% ================================================================
\section{Architecture Design}

\subsection{Persistent On-Chip State}

The foundational enabler is holding the full recurrent state---32 matrices
of $128 \times 128$ FP32 values totaling 2\,MB---permanently in BRAM
across all token steps. Unlike GPU implementations that must round-trip
state through HBM every token, our state persists on-chip via static HLS
arrays mapped to dual-port BRAM. The host invokes the kernel once per
token, passing only the $\sim$48.5\,KB of q/k/v tokens and gate values
via AXI. This satisfies constraint~\eqref{eq:c_bram} and eliminates
state I/O entirely.

\subsection{Fused Five-Phase Compute Pipeline}
\label{sec:fused}

Naive implementation of Eqs.~\eqref{eq:retrieval}-\eqref{eq:output}
requires three passes over the state matrix
(Eq.~\eqref{eq:naive_cost}). We reduce this to two by restructuring the 
algebra in state update and output steps 
% Eq.~\eqref{eq:update} 
as: 
\begin{equation}
  S_t^{\!\top} q = (g \cdot S_{t-1} + k\,\Delta v^{\!\top})^{\!\top} q
  = g \cdot S_{t-1}^{\!\top} q + (q^{\!\top}\! k)\,\Delta v
  \label{eq:fused}
\end{equation}
This allows us to compute the \emph{partial output}
$\hat{o} = g \cdot S_{t-1}^{\!\top} q$ in the same read pass as the
retrieval $r = S_{t-1}^{\!\top} k$, then correct the output as
$o = \tfrac{1}{\sqrt{d}}(\hat{o} + (q \cdot k)\,\Delta v)$ without
re-reading the updated state. Algorithm~\ref{alg:fused} shows the
resulting five-phase pipeline, which performs exactly \textbf{one read
pass and one write pass} per state matrix, reducing
$T_{\mathrm{iter}}$ from $\sim$3{,}072 to $\sim$2{,}106 cycles---a
1.46$\times$ improvement.

\begin{algorithm}[t]
\caption{Fused GDN Decode Step (Single Head)}
\label{alg:fused}
\begin{algorithmic}[1]
\REQUIRE $q, k, v \in \mathbb{R}^d$, state $S \in \mathbb{R}^{d \times d}$, gates $g, \beta$
\ENSURE output $o \in \mathbb{R}^d$, updated state $S$
\STATE $\alpha \leftarrow q^{\!\top}\! k$ \hfill\COMMENT{1: dot product}
\FOR{each row $i = 1 \ldots d$} 
  \STATE $r_i \leftarrow \sum_j S_{ji} \cdot k_j$; \quad
         $\hat{o}_i \leftarrow g \cdot \sum_j S_{ji} \cdot q_j$
\ENDFOR
\STATE $\Delta v \leftarrow \beta \cdot (v - r)$ \hfill\COMMENT{3: delta correction}
\STATE $o \leftarrow \tfrac{1}{\sqrt{d}}(\hat{o} + \alpha \cdot \Delta v)$ \hfill\COMMENT{4: output correction}
\FOR{each row $i = 1 \ldots d$}
  \STATE $S_{:,i} \leftarrow g \cdot S_{:,i} + k \cdot \Delta v_i$
\ENDFOR
\end{algorithmic}
\end{algorithm}

\noindent\textbf{Recurrence distance for II=1.} The tiled accumulation in Phase~1
has a loop-carried dependency on the partial-sum registers. With
$128/P_K = 8$ tiles per row, the recurrence distance is 8 cycles, exceeding
the FP32 fadd latency ($\sim$5 cycles), guaranteeing the HLS scheduler can
achieve II$=$1.

\subsection{GVA-Aware Paired-Head Parallelism}
\label{sec:gva}

The 2:1 GVA structure means each q/k pair serves $R{=}2$ value-heads.
We exploit this by processing both v-heads of a GVA pair simultaneously:
the q and k vectors are broadcast to both heads, while each head
maintains its own state matrix, gate values, and accumulators. This
doubles the effective compute per q/k load without duplicating q/k
storage.

More broadly, all $H_{\mathrm{iter}}$ v-heads are processed in parallel:
the innermost loops over column tiles and head indices are fully unrolled,
while the outer tile loop is pipelined at II$=$1. BRAM partitioning
ensures conflict-free access: complete partitioning along the head
dimension gives each head its own bank group, while cyclic partitioning
with factor $P_K$ along the column dimension provides $P_K$ independent
column banks per head.

\subsection{Dataflow Pipelining}
\label{sec:dataflow}

The 32 v-heads are divided into groups of $H_{\mathrm{iter}}$ heads
processed per iteration, yielding
$N_{\mathrm{iter}} = h_v / H_{\mathrm{iter}}$ iterations. Each
iteration comprises three stages:

\begin{enumerate}
\item \emph{Prepare}: copy relevant q/k/v slices and compute gates
  $g_t$, $\beta_t$ (Eqs.~\eqref{eq:gate_g}--\eqref{eq:gate_beta}).
\item \emph{Compute}: execute the fused five-phase pipeline over
  $H_{\mathrm{iter}}$ heads in parallel.
\item \emph{Store}: write $H_{\mathrm{iter}} \times d$ output elements
  to external memory via AXI.
\end{enumerate}

\noindent HLS dataflow pipelining overlaps these stages: prepare for
iteration $n{+}1$ runs concurrently with compute for iteration $n$ and
store for iteration $n{-}1$. The pipeline interval is determined by the
slowest stage---compute at $\sim$2{,}105 cycles---so the effective
iteration cost is 2{,}106 cycles regardless of $H_{\mathrm{iter}}$.
The state array is structured as a 4D tensor indexed by
$[\textit{iter}][h][i][j]$ so that each iteration accesses a disjoint
first-dimension slice, providing the HLS compiler with static proof of
no inter-iteration conflicts.

\subsection{System Overview}

% --- Figure 4: System Design ---
\begin{figure*}[t]
  \centering
  \includegraphics[width=\textwidth]{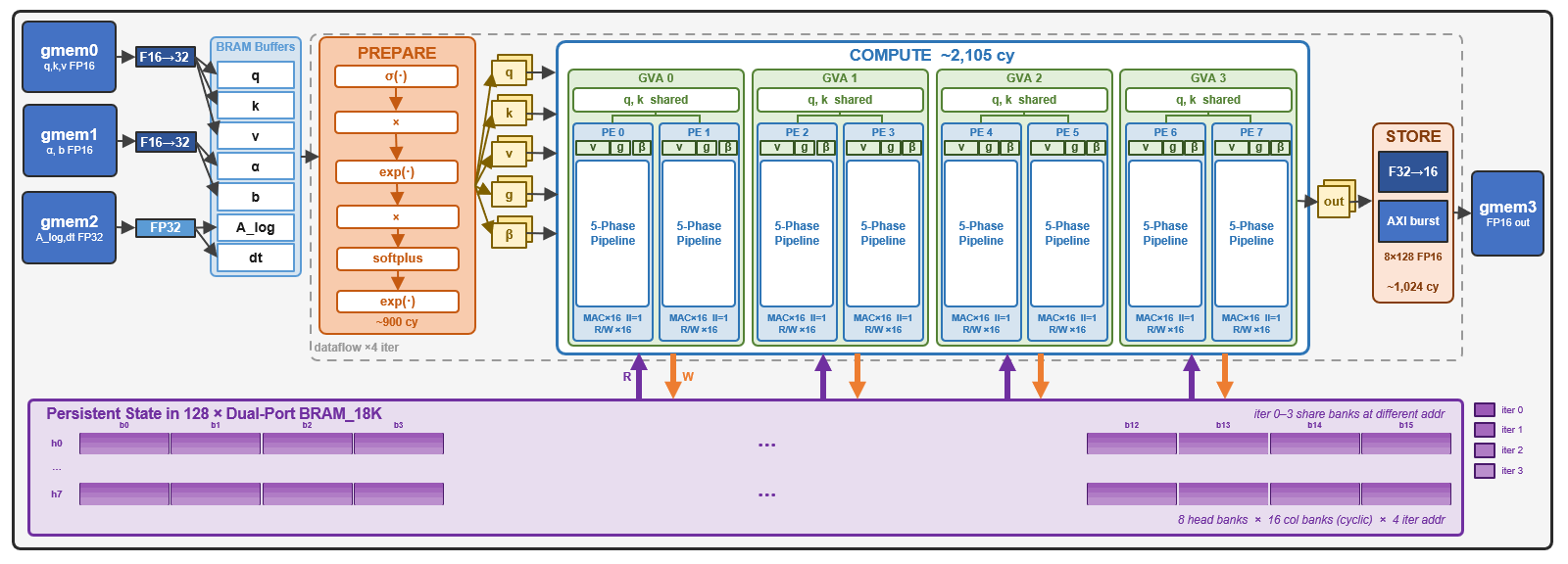}
  \caption{System architecture for $H_{\mathrm{iter}}{=}8$.
  Per-token inputs arrive via three AXI ports (left) and are buffered
  in on-chip BRAM. The prepare stage computes gates $g_t$, $\beta_t$.
  Four GVA pairs each share a q/k datapath across two PEs; each PE
  executes the fused five-phase pipeline with $P_K{=}16$ column
  parallelism (MAC$\times$16, II$=$1). The 128 dual-port BRAM arrays
  (bottom) hold the persistent 2\,MB state, partitioned by head and
  column bank, with all four iteration slices sharing the same
  physical banks at different addresses.}
  \label{fig:system}
\end{figure*}

Figure~\ref{fig:system} shows how the four techniques compose. AXI
master interfaces load the per-token inputs into on-chip buffers
($T_{\mathrm{load}}$). The dataflow loop then iterates over head groups:
each iteration's prepare stage reads from the input buffers and computes
scalar gates; the compute stage executes the fused five-phase pipeline
over $H_{\mathrm{iter}}$ heads using the persistent BRAM state; and the
store stage writes outputs back via AXI. Table~\ref{tab:configs}
summarizes the four configurations we evaluate.

\begin{table}[h]
\centering
\caption{Design-Space Configurations}
\label{tab:configs}
\begin{tabular}{lcccc}
\toprule
$H_{\mathrm{iter}}$ (v-heads/iter) & 2 & 4 & 8 & 16 \\
\midrule
$N_{\mathrm{iter}}$ (iterations) & 16 & 8 & 4 & 2 \\
GVA pairs/iter & 1 & 2 & 4 & 8 \\
Predicted $L$ @300\,MHz ($\mu$s)\textsuperscript{a} & 124.2 & 67.7 & 39.4 & 25.3 \\
\bottomrule
\multicolumn{5}{l}{\footnotesize \textsuperscript{a}Assumes constant $T_{\mathrm{iter}} \approx 2{,}106$ cycles.}
\end{tabular}
\end{table}

\noindent The predicted latency
$L = N_{\mathrm{iter}} \times 2{,}106 \times 3.33\,\text{ns}$ gives
a lower bound; measured latency is higher due to
$T_{\mathrm{load}} \approx 8{,}800$--$10{,}600$ cycles of AXI input
loading. As shown in Section~\ref{sec:results}, the constant-interval
assumption breaks down at $H_{\mathrm{iter}}{=}16$, where pipeline
inflation causes actual latency to exceed the prediction.

% ================================================================
\section{Implementation}

\subsection{Target Platform}

We target the AMD Alveo U55C accelerator card featuring 4,032 BRAM36 blocks
($\sim$17.6\,MB), 9,024 DSP48E2 slices, 2.73M flip-flops, and 1.30M
LUTs. The design is synthesized using Vitis HLS 2025.1 with a target
clock period of 3.33\,ns (300\,MHz).

\subsection{HLS Optimization Strategy}

The state matrices are mapped to dual-port BRAM, providing two independent
access ports per bank for concurrent read and write operations across
dataflow stages. Complete partitioning along the head dimension allocates
each of the $H_{\mathrm{iter}}$ heads to its own BRAM bank group,
eliminating inter-head access conflicts. Cyclic partitioning with factor
$P_K{=}16$ along the column dimension provides the parallel read/write
bandwidth required for the tiled inner loops. All tiled loops achieve an
initiation interval of one cycle, verified by HLS scheduling reports. The
top-level iteration loop uses dataflow pipelining to overlap the prepare,
compute, and store stages across successive iterations.

All configurations are validated via C/RTL
co-simulation against a golden-reference GDN decode implementation.

% ================================================================
\section{Evaluation}
\label{sec:results}

\subsection{Experimental Setup}

\noindent\textbf{Model configuration.} Qwen3-Next~\cite{qwen2025qwen3next}-style single GDN layer with
$h_q = h_k = 16$ query/key heads, $h_v = 32$ value heads, head dimension
$d = 128$. All computations in FP32.

\noindent\textbf{GPU baseline.} NVIDIA H100 PCIe (80\,GB HBM3,
2.0\,TB/s, $\sim$51\,TFLOP/s FP32) running the official NVLabs
reference implementation (PyTorch) of the GDN decode
recurrence~\cite{yang2025gated}.\footnote{\url{https://github.com/NVlabs/GatedDeltaNet}} At batch-1 the
recurrence is inherently sequential over heads: each step loads
per-token inputs, executes
Eqs.~\eqref{eq:retrieval}--\eqref{eq:output}, and stores the
output---the natural structure for a recurrence that admits no
cross-head parallelism. Latencies averaged over 1K invocations
with CUDA warm-up.

\noindent\textbf{FPGA estimates.} Post-synthesis cycle counts from Vitis HLS
$\times$ clock period. We report both 300\,MHz (synthesis target) and
250\,MHz (conservative). Vivado implementation (place-and-route) is
completed for $H_{\mathrm{iter}} \in \{2, 4\}$: $H_{\mathrm{iter}}{=}2$
achieves 263\,MHz (3.802\,ns critical path); $H_{\mathrm{iter}}{=}4$
meets post-synthesis timing (2.370\,ns critical path, identical to
$H_{\mathrm{iter}}{=}2$) but fails routing due to congestion
(88{,}725 unroutable signals), as the design exceeds a single SLR
(Figure~\ref{fig:placement}).

\begin{figure}[t]
  \centering
  \includegraphics[width=\columnwidth]{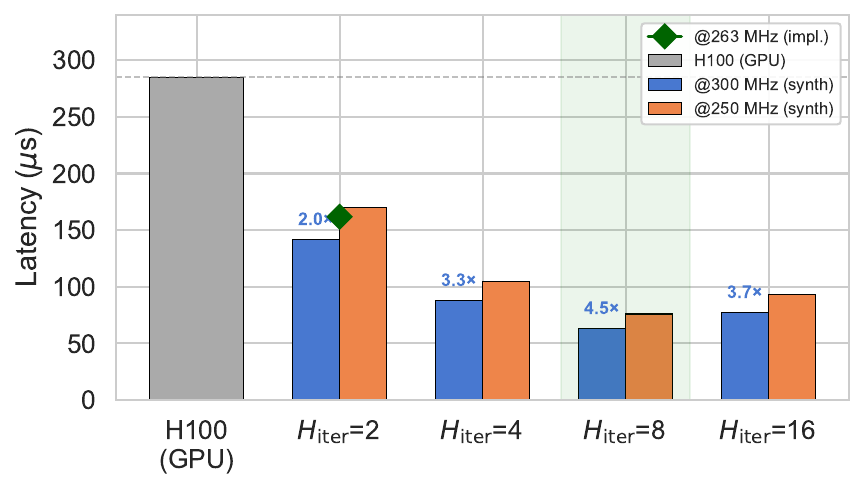}
  \caption{Per-token latency comparison for batch-1 GDN decode.
  All FPGA configurations outperform the GPU baseline.
  $H_{\mathrm{iter}}{=}8$ is optimal; $H_{\mathrm{iter}}{=}16$
  regresses due to pipeline interval inflation.}
  \label{fig:latency}
\end{figure}

\subsection{Latency Results}

\begin{table}[t]
\centering
\caption{Per-Token Decode Latency ($\mu$s) and Speedup vs.\ GPU Baseline}
\label{tab:latency}
\begin{tabular}{lcccc}
\toprule
\textbf{Platform} & \textbf{Cycles} & \textbf{@300\,MHz} & \textbf{@250\,MHz} & \textbf{Speedup} \\
\midrule
H100 (GPU) & --- & \multicolumn{2}{c}{285} & 1.0$\times$ \\
\midrule
$H_{\mathrm{iter}}{=}2$ & 42{,}538 & 141.7 & 170.2 & 2.0$\times$ \\
$H_{\mathrm{iter}}{=}4$ & 26{,}252 & 87.4 & 105.0 & 3.3$\times$ \\
$H_{\mathrm{iter}}{=}8$ & 18{,}978 & 63.2 & 75.9 & 4.5$\times$ \\
$H_{\mathrm{iter}}{=}16$ & 23{,}206 & 77.4 & 92.8 & 3.7$\times$ \\
\bottomrule
\end{tabular}
\end{table}

All four FPGA configurations outperform the H100 GPU baseline.
For $H_{\mathrm{iter}} \leq 8$, latency decreases with increasing
parallelism, following the constant dataflow interval model:
$42{,}538 \approx 16 \times 2{,}106 + 8{,}800$,
$26{,}252 \approx 8 \times 2{,}106 + 9{,}400$,
$18{,}978 \approx 4 \times 2{,}106 + 10{,}554$.
The optimal is reached at $H_{\mathrm{iter}}{=}8$ (4.5$\times$ speedup).
At $H_{\mathrm{iter}}{=}16$, however, latency \emph{increases} to
77.4\,$\mu$s (3.7$\times$) because the per-iteration interval inflates
from $\sim$2{,}106 to $\sim$6{,}300 cycles, outweighing the reduction
from 4 to 2 iterations (Section~\ref{sec:ablation}).
Even at a conservative 250\,MHz, $H_{\mathrm{iter}}{=}8$ delivers
3.8$\times$ speedup. Post-implementation timing for
$H_{\mathrm{iter}}{=}2$ confirms 263\,MHz is achievable, yielding
161.7\,$\mu$s latency per token.

\subsection{Energy Efficiency}

Post-implementation power analysis in Vivado reports a total on-chip
power of only \textbf{9.96\,W} for the placed $H_{\mathrm{iter}}{=}2$
design---comprising 6.54\,W dynamic (BRAM 1.67\,W, DSP 0.85\,W, logic
1.03\,W, clocks 1.18\,W, routing 1.82\,W) and 3.42\,W static.
Table~\ref{tab:energy} compares per-token energy.

\begin{table}[t]
\centering
\caption{Per-Token Energy Comparison}
\label{tab:energy}
\begin{tabular}{lccc}
\toprule
\textbf{Platform} & \textbf{Power} & \textbf{Latency} & \textbf{Energy/token} \\
\midrule
H100 PCIe (GPU) & 350\,W\textsuperscript{a} & 285\,$\mu$s & 99.8\,mJ \\
\midrule
$H_{\mathrm{iter}}{=}2$ (impl.) & 9.96\,W\textsuperscript{b} & 161.7\,$\mu$s & 1.61\,mJ \\
$H_{\mathrm{iter}}{=}4$ & $\leq$150\,W\textsuperscript{a} & 87.4\,$\mu$s & $\leq$13.1\,mJ \\
$H_{\mathrm{iter}}{=}8$ & $\leq$150\,W\textsuperscript{a} & 63.2\,$\mu$s & $\leq$9.5\,mJ \\
$H_{\mathrm{iter}}{=}16$ & $\leq$150\,W\textsuperscript{a} & 77.4\,$\mu$s & $\leq$11.6\,mJ \\
\bottomrule
\multicolumn{4}{l}{\footnotesize \textsuperscript{a}Board-level TDP~\cite{nvidia_h100,xilinx_u55c} (conservative upper bound).} \\
\multicolumn{4}{l}{\footnotesize \textsuperscript{b}Post-implementation Vivado power estimate at 263\,MHz.}
\end{tabular}
\end{table}

\noindent At 1.61\,mJ per token, the implemented $H_{\mathrm{iter}}{=}2$
configuration is \textbf{62$\times$} more energy-efficient than the GPU
baseline. This dramatic gap arises from two compounding factors:
(1)~the FPGA eliminates the HBM state round-trip, completing the
workload faster; and (2)~actual on-chip power at 12\% resource
utilization is far below the 150\,W board
TDP~\cite{xilinx_u55c}. Even assuming the full 150\,W board TDP,
energy efficiency ranges from 7.6$\times$ to 10.5$\times$.

\begin{figure}[t]
  \centering
  \includegraphics[width=\columnwidth]{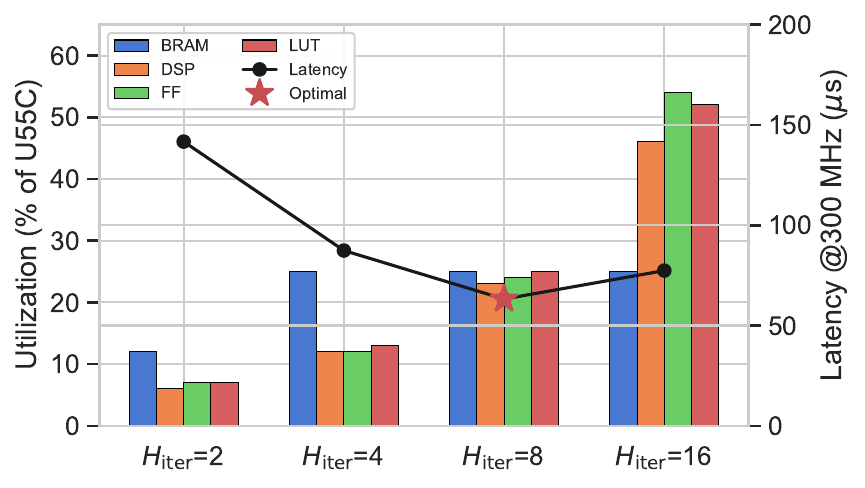}
  \caption{FPGA resource utilization and latency across configurations.
  BRAM plateaus at 25\%; DSP, FF, and LUT scale with $H_{\mathrm{iter}}$.
  Latency (right axis) reaches a minimum at $H_{\mathrm{iter}}{=}8$;
  $H_{\mathrm{iter}}{=}16$ increases latency due to pipeline interval
  inflation despite fewer iterations.}
  \label{fig:resources}
\end{figure}

\subsection{Resource Utilization}

\begin{table}[t]
\centering
\caption{FPGA Resource Utilization (Alveo U55C)}
\label{tab:resources}
\begin{tabular}{lcccccc}
\toprule
& \multicolumn{2}{c}{\textbf{BRAM\textsuperscript{a}}} & \multicolumn{2}{c}{\textbf{DSP}} & \textbf{FF} & \textbf{LUT} \\
& Used & \% & Used & \% & \% & \% \\
\midrule
$H_{\mathrm{iter}}{=}2$ & 523 & 12 & 570 & 6 & 7 & 7 \\
$H_{\mathrm{iter}}{=}4$ & 1{,}035 & 25 & 1{,}090 & 12 & 12 & 13 \\
$H_{\mathrm{iter}}{=}8$ & 1{,}035 & 25 & 2{,}130 & 23 & 24 & 25 \\
$H_{\mathrm{iter}}{=}16$ & 1{,}035 & 25 & 4{,}162 & 46 & 54 & 52 \\
\bottomrule
\multicolumn{7}{l}{\footnotesize \textsuperscript{a}BRAM\_18K count; 4{,}032 available.}
\end{tabular}
\end{table}

BRAM usage is dominated by the persistent state storage (32 matrices of
$128 \times 128$ FP32 values) and plateaus at 25\% once the head dimension is
fully partitioned ($H_{\mathrm{iter}} \geq 4$). DSP, FF, and LUT scale
approximately linearly with $H_{\mathrm{iter}}$ as each additional head
requires its own set of multiply-accumulate units and state-access ports.
At $H_{\mathrm{iter}}{=}8$, all resource types stay at or below 25\%.
Pushing to $H_{\mathrm{iter}}{=}16$ roughly doubles compute resources
(46\% DSP, 54\% FF, 52\% LUT) while actually \emph{increasing} latency,
making $H_{\mathrm{iter}}{=}8$ the optimal configuration.

\subsection{Ablation Analysis}
\label{sec:ablation}

For $H_{\mathrm{iter}} \in \{2, 4, 8\}$, the compute stage dominates
each dataflow iteration at a near-constant $\sim$2{,}105 cycles.
Total latency follows the model
$L \approx N_{\mathrm{iter}} \times 2{,}106 + L_{\mathrm{load}}$,
where $L_{\mathrm{load}} \approx 8{,}800$--$10{,}600$ cycles.

At $H_{\mathrm{iter}}{=}16$, this model breaks down. The per-iteration
interval inflates from $\sim$2{,}106 to $\sim$6{,}300 cycles because
the wider datapath (16 heads unrolled) increases pipeline depth and
routing pressure, raising the effective initiation interval of the
tiled loops. The result is that $H_{\mathrm{iter}}{=}16$ at 23{,}206
total cycles is \emph{slower} than $H_{\mathrm{iter}}{=}8$ at 18{,}978
despite having half the iterations. This identifies
$H_{\mathrm{iter}}{=}8$ as the optimal configuration: it
minimizes latency at 4.5$\times$ speedup while staying under 25\%
utilization for all resource types.

\begin{figure}[t]
  \centering
  \includegraphics[width=0.48\columnwidth]{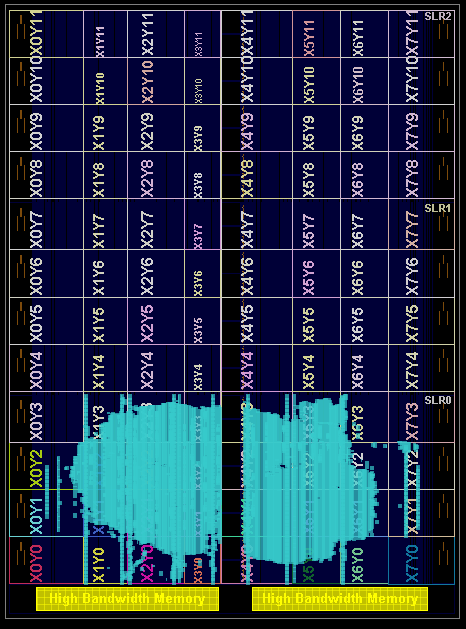}\hfill
  \includegraphics[width=0.48\columnwidth]{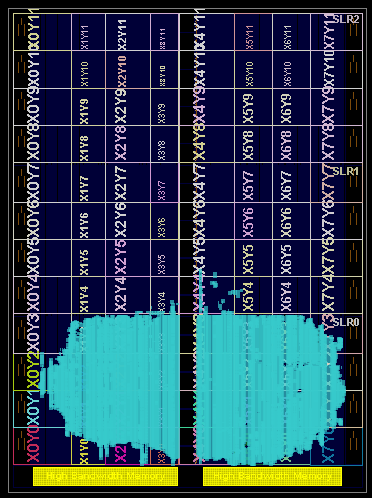}
  \caption{Post-implementation placement on Alveo U55C.
  \textbf{Left:} $H_{\mathrm{iter}}{=}2$ fits comfortably within SLR0
  (12\% BRAM, 6\% DSP; 9.96\,W on-chip power, 263\,MHz).
  \textbf{Right:} $H_{\mathrm{iter}}{=}4$ saturates SLR0 and spills
  into SLR1 (77\% SLR-level BRAM, 39\% LUT), causing routing failure
  (88{,}725 unroutable signals) despite meeting post-synthesis timing
  at 2.370\,ns.}
  \label{fig:placement}
\end{figure}

\noindent\textbf{Physical scaling wall.}
Post-implementation results reinforce the synthesis trends
(Figure~\ref{fig:placement}). $H_{\mathrm{iter}}{=}2$ places and routes
within a single SLR at 263\,MHz. $H_{\mathrm{iter}}{=}4$ passes
post-synthesis timing with an identical 2.370\,ns critical path but
fails routing: at 77\% SLR-level BRAM and 39\% LUT, the design
saturates SLR0 and spills across SLR boundaries, causing 88{,}725
unroutable signals.  This confirms that
$H_{\mathrm{iter}} \in \{8, 16\}$ would require explicit multi-SLR
floorplanning or a larger device, and that $H_{\mathrm{iter}}{=}2$
represents the most implementable design point at the cost of higher
latency.

% ================================================================
\section{Related Work}

FPGA acceleration of Transformer-based LLMs has been explored
extensively~\cite{hong2022dfx,chen2024spatial,zeng2024flightllm,zhang2026flexllm,he2025lutllm,guan2017fpga,han2017ese}.
DFX~\cite{hong2022dfx} demonstrated multi-FPGA dataflow for GPT-2
decode; Chen et al.~\cite{chen2024spatial} provided an analytical
framework for spatial LLM acceleration;
FlightLLM~\cite{zeng2024flightllm} and FlexLLM~\cite{zhang2026flexllm}
target sparse Transformer and composable HLS design respectively; and
LUT-LLM~\cite{he2025lutllm} replaces multiply-accumulate with
memory-based table lookups.
Since GDN is a strict superset of Mamba-2 (adding the delta rule and
gating), FPGA work on Mamba is 
relevant: MARCA~\cite{marca2024} introduced a reconfigurable PE array
for Mamba's mixed linear and element-wise operations,
LightMamba~\cite{lightmamba2025} co-designs quantization with FPGA
dataflow for efficient Mamba inference, and
SpecMamba~\cite{specmamba2025} combines speculative decoding with
customized Mamba dataflow on Versal FPGAs.
% However, all of the above target either standard softmax attention,
% MLP layers, or Mamba's simpler diagonal-state SSM recurrence; none
% address the $d \times d$ matrix-state recurrence of GDN or exploit
% persistent on-chip state to eliminate the HBM round-trip.
However, all of the above target either standard softmax attention,                                                                                    
  MLP layers, or Mamba's simpler diagonal-state SSM recurrence.                                                                                          
  To the best of our knowledge, we are the first to accelerate the                                                                                       
  $d \times d$ matrix-state recurrence of GDN at full precision on a                                                                                     
  datacenter FPGA, targeting low-energy, production-scale LLM inference.

\section{Conclusion}

% --- Figure: Placement comparison ---

% We presented an FPGA accelerator for Gated DeltaNet decode that eliminates
% the HBM state-transfer bottleneck limiting GPU performance at batch size~1.
% By holding the full recurrent state persistently in on-chip BRAM, our design
% converts the workload from memory-bound to compute-bound. A fused
% five-phase pipeline, GVA-aware head pairing, and dataflow pipelining yield
% 4.5$\times$ lower latency and over 60$\times$ energy efficiency per
% token over the GPU baseline on H100~PCIe, with post-implementation
% power measured at under 10\,W.
% We presented an FPGA accelerator for Gated DeltaNet decode that eliminates                                                                             
%   the HBM state-transfer bottleneck limiting batch-1 GPU performance,
%   achieving 4.5$\times$ lower latency and over 60$\times$ energy efficiency                                                                              
%   at under 10\,W. Unlike attention, whose KV cache grows with
%   sequence length, GDN's fixed-size state fits entirely in BRAM---making
%   subquadratic architectures a natural match for FPGA. Extending this to
%   prefill, mixed-precision quantization, sparse MoE routing, and
%   co-acceleration of the remaining softmax layers would enable end-to-end
%   hybrid LLM inference on a single datacenter FPGA.

% As hybrid LLM architectures increasingly adopt linear attention layers,
% FPGAs' ability to maintain persistent on-chip state positions them as
% efficient co-processors for the memory-bound decode primitives that
% dominate serving latency.

We presented an FPGA accelerator for Gated DeltaNet decode that eliminates
  the HBM state-transfer bottleneck limiting batch-1 GPU performance,                                                                                    
  achieving 4.5$\times$ lower latency and over 60$\times$ greater energy
  efficiency at under 10\,W. As subquadratic models with fixed-size state increasingly                                                                              
  replace attention layers in production LLMs, memory-boundedness becomes the
  dominant inference bottleneck. Future work will extend the persistent-state
  datapath to support prefill, mixed-precision quantization, sparse MoE
  routing, and co-acceleration of the remaining softmax layers to enable large-scale
  hybrid LLM inference on a single datacenter FPGA.

% ================================================================
\section*{Acknowledgment}
This work is supported by the NSF under grant numbers CNS-2009057, CSSI-2311870, CCF-1919289, and SPX-2333009.

\bibliographystyle{IEEEtran}
\bibliography{main}

\end{document}